\documentclass{article}
\usepackage{amssymb}
\usepackage{amsmath}
\usepackage{graphicx}

\input{tcilatex}
\begin{document}

\title{Quantum Games and Quantum Discord}
\author{Ahmad Nawaz\thanks{%
email: ahmad@ele.qau.edu.pk} \ and A. H. Toor\thanks{
\ \ \ \ \ \ \ \ \ ahtoor@qau.edu.pk}$^{\dag }$ \\
National Centre for Physics, Quaid-i-Azam University Campus, \\
Islamabad, Pakistan\\
\dag\ Department of Physics, Quaid-i-Azam University, \\
Islamabad 45320, Pakistan}
\maketitle

\begin{abstract}
We quantize prisoners dilemma and chicken game by our generalized
quantization scheme to explore the role of quantum discord in quantum games.
In order to establish this connection we use Werner-like state as an initial
state of the game. In this quantization scheme measurement can be performed
in entangled as well as in product basis.\emph{\ }For the measurement in
entangled basis the dilemma in both the games can be resolved by separable
states with non-zero quantum discord.\textrm{\ }Similarly for \textrm{%
product basis measurement the payoffs are quantum mechanical only for
nonzero values of quantum discord.}
\end{abstract}

\section{Introduction}

Entanglement is a key resource in quantum information theory. If used as a
resource it can perform numerous tasks which seem rather impossible for
classical resources and shared randomness. Quantum game theory is no
exception in this regard where entanglement plays vital role in the solution
of games. The first step to this direction was by Eisert \textit{et al}. 
\cite{eisert} who introduced an elegant scheme for the quantization of
prisoner dilemma (PD). In this scheme the strategy space of the players is a
two parameter set of $2\times 2$ unitary matrices. Starting with maximally
entangled initial state the authors showed that for a suitable quantum
strategy the dilemma disappeared. They also pointed out a quantum strategy
which always won over all the classical strategies. Later on Marinatto and
Weber \cite{marinatto} introduced another interesting and simple scheme for
the analysis of non-zero sum classical games in quantum domain. They gave
Hilbert structure to the strategic spaces of the players. They also used
maximally entangled initial state and allowed the players to play their
tactics by applying probabilistic choice of unitary operators. They applied
their scheme to an interesting game of Battle of Sexes and found out the
strategy for which both the players can achieve equal payoffs. Both these
quantization\emph{\ }schemes gave interesting results for various quantum
analogues of classical games \cite%
{flitney,azhar,azhar1,azhar2,jiang,rosero,nawazbattle}. In all these cases
entanglement played a crucial role. However, it has recently been shown that
\ all the non classical properties of quantum correlations can not be
analyzed by entanglement only \cite{henderson,harold}. \ Instead of
entanglement quantum discord (QD) is introduced as a feature of quantumness
of correlations that remains there even for separable states. \ Since then
various aspects of QD has been studied \cite{piani,Ferraro,Werlang,Datta}.

In this paper we explore the role of QD in quantization of games. To
elaborate this role we start our game with Werner-like states which have
very interesting proprties. These states are linear combination of a
maximally entangled and a maximally mixed state \cite{werner,munro,ghosh,wei}%
. Their entanglement and nonlocality depends upon a parameter $0$ $\leq
p\leq 1$ that parameterizes them. For $0<p\leq \frac{1}{3}$ they are
separable,\ for $\frac{1}{3}$ $<p\leq \frac{1}{\sqrt{2}}$ entangled but not
nonlocal and for the range $\frac{1}{\sqrt{2}}<p<1$ they become inseparable
and nonlocal \cite{werner}. In the recent years these states have been
investigated from different perspective like entanglement teleportation via
Werner states \cite{lee}, entanglement of Werner derivatives \textrm{\cite%
{hiroshima}}, Bell violation and entanglement \ of Werner states of two
qubits in independent decay channels \cite{miranowicz} and their application
in ancilla assisted process tomography \cite{altepeter}. It has also been
reported that despite being nonlocal, for certain range of parameter $p,$
when shared between two parties these states are a powerful\ resource in
comparison to classical randomness \textrm{\cite{preskil}}. In Ref. \cite%
{harold} the QD for these states has been found and shown positive for all $%
p>0.$ This behavior of these states is in contrast with their well known
separability at $p\leq \frac{1}{3}$. Here we quantize PD and chicken game
(CG) using our generalized quantization scheme taking Werner-like state as
an initial quantum state \cite{nawaz}. In this scheme measurement can be
performed in both entangled and product basis.\emph{\ }For the measurement
in entangled basis our results show that the strategy pair $\left(
Q,Q\right) $ remains Nash equilibrium for both these games as in the
quantization scheme of Eisert et al. \cite{eisert} for all values of $p>0$.
It is interesting to note that for $p\leq \frac{1}{3}$ unentangled quantum
state with non zero QD is capable of resolving dilemmas in PD and CG. This
shows that QD also has a crucial role in quantum games. For $p=0$ when QD\
becomes zero then the payoffs become constant and independent of the players
strategies.\textrm{\ For the second case when the measurement is performed
in product basis the payoffs remain quantum mechanical as in Marinatto and
Weber scheme only for }$p>0$ \textrm{\ i.e. for nonzero values of quantum
discord \cite{marinatto,ma}.}

This paper is organized as follows; in sections (\ref{prisoner}) and (\ref%
{chicken}) we give a brief introduction to the classical versions of PD and
CG respectively, section (\ref{quantum}) deals with quantum discord \cite%
{harold} and its role in quantum games and section (\ref{conclusion})
concludes the results.

\section{\label{prisoner}Prisoners' Dilemma\textbf{\ }}

This game starts with a story of two suspects, say Alice and Bob, who have
committed a crime together. Now they are being interrogated in a separate
cell. The two possible moves for each player are to cooperate ($C$) or to
defect ($D$) without any communication between them according to the
following payoff matrix%
\begin{equation}
\text{{\large Alice }}%
\begin{array}{c}
C \\ 
D%
\end{array}%
\overset{}{\overset{%
\begin{array}{c}
\text{{\large Bob}} \\ 
\begin{array}{cc}
C\text{ \ \ \ \ } & D%
\end{array}%
\end{array}%
}{\left[ 
\begin{array}{cc}
\left( 3,3\right) & \left( 0,5\right) \\ 
\left( 5,0\right) & \left( 1,1\right)%
\end{array}%
\right] }}.  \label{matrix-prisoner}
\end{equation}%
\emph{\ }It is clear from the above payoff matrix that $D$\ is the dominant
strategy for both players. Therefore, rational reasoning forces each player
to play $D$. Thus ($D,D$) results as the Nash equilibrium of this game with
payoffs $(1,1),$ which is not Pareto Optimal. However, it was possible for
the players to get higher payoffs if they would have played $C$\ instead of $%
D$. This is the origin of dilemma in this game.

\section{\label{chicken}Chicken Game}

The payoff matrix for this game is%
\begin{equation}
\text{{\large Alice }}%
\begin{array}{c}
C \\ 
D%
\end{array}%
\overset{}{\overset{%
\begin{array}{c}
\text{{\large Bob}} \\ 
\begin{array}{cc}
C\text{ \ \ \ \ } & D%
\end{array}%
\end{array}%
}{\left[ 
\begin{array}{cc}
\left( 3,3\right) & \left( 1,4\right) \\ 
\left( 4,1\right) & \left( 0,0\right)%
\end{array}%
\right] }}.  \label{matrix chicken}
\end{equation}%
In this game two players drove their cars straight towards each other. The
first to swerve to avoid a collision is the loser (chicken) and the one who
keeps on driving straight is the winner. There is no dominant strategy in
this game. There are two Nash equilibria $\left( C,D\right) $ and $\left(
D,C\right) ,$ the former is preferred by Bob and the latter is preferred by
Alice. The dilemma of this game is that the Pareto Optimal strategy $\left(
C,C\right) $ is not NE.

\section{\label{quantum}Quantum Discord and Quantum Games}

The Shannon entropy of a discrete variable $X$ with discrete probability
distribution $p_{x}$ is defined as 
\begin{equation}
H\left( X\right) =-\underset{x}{\tsum }p_{x}\log p_{x}.
\end{equation}%
The conditional entropy of $X$ given $Y$ is the measure of the amount of
uncertainty about $X$ given the value of $Y.$ Mathematically it is written
as 
\begin{equation}
H\left( X\mid Y\right) =-\underset{x,y}{\tsum }p\left( x,y\right) \log
p\left( x\mid y\right)
\end{equation}%
where $p\left( x,y\right) $ is the joint probability distribution of the
random variable $X$ and $Y$ and $p\left( x\mid y\right) \ $ is the
conditional probability of the occurrence of $X$ when $Y$ has already
occurred$.$ The correlation between two random variables $X,$ $Y$ with
probability distributions $p_{x}$ and $p_{y}$ respectively is called mutual
information. Mathematically it takes the form 
\begin{equation}
I\left( X:Y\right) =H\left( X\right) +H\left( Y\right) -H\left( X,Y\right) .
\label{mutual01}
\end{equation}%
where $H\left( X,Y\right) $ is joint entropy that measures the average
uncertainty of the pair $\left( X,Y\right) .$ Mutual information given in
Eq. (\ref{mutual01}) can also be written as%
\begin{equation}
J\left( X:Y\right) =H\left( X\right) -H\left( X\mid Y\right) =H\left(
Y\right) -H\left( B\mid Y\right) .  \label{mutual02}
\end{equation}%
Quantum analogue of Shannon entropy for a quantum system in state $\rho $ is
von Neumann entropy which is given as 
\begin{equation}
S\left( \rho \right) =-\text{Tr}\left( \rho \log \rho \right)
\end{equation}%
that leads to the mutual information relation for state $\rho _{XY}$ to be
written as 
\begin{equation}
I\left( \rho _{XY}\right) =S\left( \rho _{X}\right) +S\left( \rho
_{Y}\right) -S\left( \rho _{XY}\right)  \label{mutualq01}
\end{equation}%
But similar generalization for Eq. (\ref{mutual02}) in quantum domain is not
straight forward. This is because that quantum conditional entropy $S\left(
\rho _{X}\mid \rho _{Y}\right) $ requires to specify the state of system $%
\rho _{X}$ given the state of $\rho _{Y}$. \ This statement in quantum
mechanics is ambiguous until measurement operators $\Pi _{i}^{Y}$ for state $%
\rho _{Y}$ are defined. If the measurement is performed using operators $\Pi
_{i}^{Y}$ then Eq. (\ref{mutual02}) in quantum domain takes the form 
\begin{equation}
J\left( \rho _{XY}\right) =S\left( \rho _{X}\right) -S\left( \rho _{X}\mid
\Pi _{i}^{Y}\right)  \label{mutualq02}
\end{equation}%
where 
\begin{equation}
S\left( \rho _{X}\mid \Pi _{i}^{Y}\right) =\underset{i}{\tsum p_{i}}S\left(
\rho _{X\mid \Pi _{i}^{Y}}\right) .
\end{equation}%
Quantum discord is defined as \cite{harold}%
\begin{equation}
D\left( X:Y\right) =I\left( \rho _{XY}\right) -J\left( \rho _{XY}\right)
\end{equation}%
that with the help of Eqs. (\ref{mutualq01}) and (\ref{mutualq02}) becomes 
\begin{equation}
D\left( X:Y\right) =S\left( \rho _{X}\right) -S\left( \rho _{XY}\right)
+S\left( \rho _{X}\mid \Pi _{i}^{Y}\right) .
\end{equation}%
For two qubit Werner like state \cite{werner} of the form 
\begin{equation}
\rho _{in}=p\left\vert \phi ^{+}\right\rangle \left\langle \phi
^{+}\right\vert +\frac{\left( 1-p\right) }{4}I\otimes I  \label{state in}
\end{equation}%
where $\left\vert \phi ^{+}\right\rangle $ $=\frac{\left\vert
00\right\rangle +\left\vert 11\right\rangle }{\sqrt{2}}$ is standard Bell
state, the quantum discord is shown below \cite{harold}. 
\begin{figure}[th]
\centering
\includegraphics[scale=1.1]{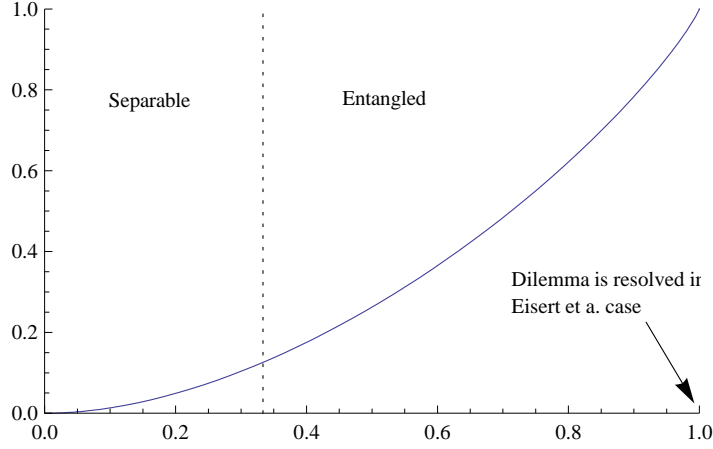}
\caption{Quantum Discord for Werner states}
\label{fig:image}
\end{figure}
It is clear from this graph that QD is greater than zero even for separable
states i.e. for $p<\frac{1}{3}.$

Next we quantize PD and CG\ \ taking Werner like state (\ref{state in}) as
an initial quantum state of the game with a payoff matrix of the form

\begin{equation}
\text{{\large Alice }}%
\begin{array}{c}
A_{1} \\ 
A_{2}%
\end{array}%
\overset{}{\overset{%
\begin{array}{c}
\text{{\large Bob}} \\ 
\begin{array}{cc}
B_{1}\text{\ \ \ \ } & B_{1}%
\end{array}%
\end{array}%
}{\left[ 
\begin{array}{cc}
\left( \$_{00}^{A},\$_{00}^{B}\right) & \left( \$_{01}^{A},\$_{01}^{B}\right)
\\ 
\left( \$_{10}^{A},\$_{10}^{B}\right) & \left( \$_{11}^{A},\$_{11}^{B}\right)%
\end{array}%
\right] }}  \label{payoff-matrix-general}
\end{equation}%
The strategy of each of the\ players is represented by the unitary operator $%
U_{i}$\ given as\emph{\ } 
\begin{equation}
U_{i}=\cos \frac{\theta _{i}}{2}R_{i}+\sin \frac{\theta _{i}}{2}C_{i},\text{
\ \ \ }  \label{strategy}
\end{equation}%
where $i=1$\ or $2$\ and $R_{i}$, $C_{i}$\emph{\ }are the unitary operators
defined as%
\begin{align}
R_{i}\left\vert 0\right\rangle & =e^{i\phi _{i}}\left\vert 0\right\rangle ,%
\text{ \ \ }R_{i}\left\vert 1\right\rangle =e^{-i\phi _{i}}\left\vert
1\right\rangle ,  \notag \\
C_{i}\left\vert 0\right\rangle & =-\left\vert 1\right\rangle ,\text{ \ \ \ \
\ }C_{i}\left\vert 1\right\rangle =\left\vert 0\right\rangle .
\label{operators}
\end{align}%
Here we restrict our treatment to two parameter set of strategies for
mathematical simplicity in accordance with Ref. \cite{eisert}.\emph{\ }After
the application of the strategies, the initial state given by Eq. (\ref%
{state in}) transforms into 
\begin{equation}
\rho _{f}=(U_{1}\otimes U_{2})\rho _{in}(U_{1}\otimes U_{2})^{\dagger }.
\label{final}
\end{equation}%
The payoff operators for Alice and Bob are

\begin{align}
P^{A}&
=\$_{00}^{A}P_{00}+\$_{11}^{A}P_{11}+\$_{01}^{A}P_{01}+\$_{10}^{A}P_{10}, 
\notag \\
P^{B}&
=\$_{00}^{B}P_{00}+\$_{11}^{B}P_{11}+\$_{01}^{B}P_{01}+\$_{10}^{B}P_{10},
\label{pay-operator}
\end{align}%
where 
\begin{subequations}
\begin{align}
P_{00}& =\left\vert \psi _{00}\right\rangle \left\langle \psi
_{00}\right\vert \text{, \ }\left\vert \psi _{00}\right\rangle =\cos \frac{%
\delta }{2}\left\vert 00\right\rangle +i\sin \frac{\delta }{2}\left\vert
11\right\rangle ,  \label{oper 1} \\
P_{11}& =\left\vert \psi _{11}\right\rangle \left\langle \psi
_{11}\right\vert ,\text{ \ }\left\vert \psi _{11}\right\rangle =\cos \frac{%
\delta }{2}\left\vert 11\right\rangle +i\sin \frac{\delta }{2}\left\vert
00\right\rangle ,  \label{oper 2} \\
P_{10}& =\left\vert \psi _{10}\right\rangle \left\langle \psi
_{10}\right\vert \text{, \ }\left\vert \psi _{10}\right\rangle =\cos \frac{%
\delta }{2}\left\vert 10\right\rangle -i\sin \frac{\delta }{2}\left\vert
01\right\rangle ,  \label{oper 3} \\
P_{01}& =\left\vert \psi _{01}\right\rangle \left\langle \psi
_{01}\right\vert \text{, \ }\left\vert \psi _{01}\right\rangle =\cos \frac{%
\delta }{2}\left\vert 01\right\rangle -i\sin \frac{\delta }{2}\left\vert
10\right\rangle ,  \label{oper 4}
\end{align}%
with\emph{\ }$\delta \in \left[ 0,\frac{\pi }{2}\right] $ being the
entanglement of the measurement basis. Above payoff operators reduce to that
of Eisert's scheme for $\delta $ equal to $\gamma ,$ which represents the
entanglement of the initial state \cite{eisert}. For $\delta =0$ above
operators transform into that of Marinatto and Weber's scheme \cite%
{marinatto}. The payoffs for the players are calculated as 
\end{subequations}
\begin{eqnarray}
\$_{A}(\theta _{1},\phi _{1},\theta _{2},\phi _{2}) &=&\text{Tr}(P^{A}\rho
_{f})\text{,}  \notag \\
\$_{B}(\theta _{1},\phi _{1},\theta _{2},\phi _{2}) &=&\text{Tr}(P^{B}\rho
_{f}),  \label{payoff-generalized}
\end{eqnarray}%
where Tr represents the trace of a\emph{\ }matrix. Using Eqs. (\ref{final}),
(\ref{pay-operator}) and (\ref{payoff-generalized}) the payoffs for players $%
j=A,B$ are obtained as%
\begin{eqnarray}
\$_{j}(\theta _{1},\phi _{1},\theta _{2},\phi _{2}) &=&\$_{00}^{j}\text{Tr}%
(P_{00}\rho _{f})+\$_{01}^{j}\text{Tr}(P_{01}\rho _{f})+\$_{10}^{j}\text{Tr}%
(P_{10}\rho _{f})+\$_{11}^{j}\text{Tr}(P_{11}\rho _{f})  \notag \\
&&  \label{payoffs}
\end{eqnarray}%
where we have defined 
\begin{subequations}
\label{a}
\begin{eqnarray}
\text{Tr}(P_{00}\rho _{f}) &=&p\left[ \left\{ 1-\sin ^{2}\left( \phi
_{1}+\phi _{2}\right) \sin \delta \right\} \cos ^{2}\frac{\theta _{1}}{2}%
\cos ^{2}\frac{\theta _{2}}{2}+\right.  \notag \\
&&\frac{\left( \sin \delta -1\right) }{2}\left\{ \cos ^{2}\frac{\theta _{1}}{%
2}+\cos ^{2}\frac{\theta _{2}}{2}-\frac{1}{2}\sin \theta _{1}\sin \theta
_{2}\sin \left( \phi _{1}+\phi _{2}\right) \right\} -  \notag \\
&&\left. \frac{\sin \delta }{2}\right] +\frac{1+p}{4}  \label{tr}
\end{eqnarray}%
\end{subequations}
\begin{eqnarray}
\text{Tr}(P_{01}\rho _{f}) &=&p\left[ \frac{1+\cos 2\phi _{1}\sin \delta }{2}%
\cos ^{2}\frac{\theta _{1}}{2}\sin ^{2}\frac{\theta _{2}}{2}+\frac{1-\cos
2\phi _{2}\sin \delta }{2}\sin ^{2}\frac{\theta _{1}}{2}\cos ^{2}\frac{%
\theta _{2}}{2}\right. +  \notag \\
&&\left. \frac{\left( -1+\sin \delta \right) \sin \phi _{1}\cos \phi
_{2}-\left( 1+\sin \delta \right) \cos \phi _{1}\sin \phi _{2}}{4}\sin
\theta _{1}\sin \theta _{2}\right] +\frac{1-p}{4}  \notag \\
&&  \label{trb}
\end{eqnarray}%
\begin{eqnarray}
\text{Tr}(P_{10}\rho _{f}) &=&p\left[ \frac{1-\cos 2\phi _{1}\sin \delta }{2}%
\cos ^{2}\frac{\theta _{1}}{2}\sin ^{2}\frac{\theta _{2}}{2}+\frac{1+\cos
2\phi _{2}\sin \delta }{2}\sin ^{2}\frac{\theta _{1}}{2}\cos ^{2}\frac{%
\theta _{2}}{2}\right. -  \notag \\
&&\left. \frac{\left( 1+\sin \delta \right) \sin \phi _{1}\cos \phi
_{2}+\left( 1-\sin \delta \right) \cos \phi _{1}\sin \phi _{2}}{4}\sin
\theta _{1}\sin \theta _{2}\right] +\frac{1-p}{4}  \notag \\
&&  \label{trc}
\end{eqnarray}%
\begin{eqnarray}
\text{Tr}(P_{11}\rho _{f}) &=&p\left[ \left\{ 1-\cos ^{2}\left( \phi
_{1}+\phi _{2}\right) \sin \delta \right\} \cos ^{2}\frac{\theta _{1}}{2}%
\cos ^{2}\frac{\theta _{2}}{2}+\right.  \notag \\
&&\left. \frac{\left( \sin \delta +1\right) }{2}\left\{ \sin ^{2}\frac{%
\theta _{1}}{2}\sin ^{2}\frac{\theta _{2}}{2}+\frac{1}{2}\sin \theta
_{1}\sin \theta _{2}\sin \left( \phi _{1}+\phi _{2}\right) \right\} \right] +
\notag \\
&&+\frac{1-p}{4}  \label{trd}
\end{eqnarray}%
In the framework of our generalized quantization scheme \cite{nawaz}
measurement can be performed either using entangled basis $\left( \delta =%
\frac{\pi }{2}\right) $ or product basis $\left( \delta =0\right) $. For the
measurement in entangled basis with the help of Eq. (\ref{payoffs}) the
payoffs for players become 
\begin{eqnarray}
\$_{j}(\theta _{1},\phi _{1},\theta _{2},\phi _{2}) &=&p\left[
\$_{00}^{j}\left( \cos ^{2}\left( \phi _{1}+\phi _{2}\right) \cos ^{2}\frac{%
\theta _{1}}{2}\cos ^{2}\frac{\theta _{2}}{2}\right) \right. +  \notag \\
&&\$_{01}^{j}\left( \cos \phi _{1}\cos \frac{\theta _{1}}{2}\sin \frac{%
\theta _{2}}{2}-\sin \phi _{2}\sin \frac{\theta _{1}}{2}\cos \frac{\theta
_{2}}{2}\right) ^{2}+  \notag \\
&&\$_{10}^{j}\left( \sin \phi _{1}\cos \frac{\theta _{1}}{2}\sin \frac{%
\theta _{2}}{2}-\cos \phi _{2}\sin \frac{\theta _{1}}{2}\cos \frac{\theta
_{2}}{2}\right) ^{2}+  \notag \\
&&\left. \$_{11}^{j}\left( \cos \frac{\theta _{1}}{2}\cos \frac{\theta _{2}}{%
2}\sin \left( \phi _{1}+\phi _{2}\right) +\sin \frac{\theta _{1}}{2}\sin 
\frac{\theta _{2}}{2}\right) ^{2}\right]  \notag \\
&&+\frac{\left( 1-p\right) }{4}\left(
\$_{00}^{j}+\$_{01}^{j}+\$_{10}^{j}+\$_{11}^{j}\right)
\label{payoff-delta-Pi/2}
\end{eqnarray}%
For PD with the payoff matrix (\ref{matrix-prisoner}) the above equation
reduce to 
\begin{eqnarray}
\$_{A}(\theta _{1},\phi _{1},\theta _{2},\phi _{2}) &=&p\left[ 3\left( \cos
^{2}\left( \phi _{1}+\phi _{2}\right) \cos ^{2}\frac{\theta _{1}}{2}\cos ^{2}%
\frac{\theta _{2}}{2}\right) \right. +  \notag \\
&&5\left( \sin \phi _{1}\cos \frac{\theta _{1}}{2}\sin \frac{\theta _{2}}{2}%
-\cos \phi _{2}\sin \frac{\theta _{1}}{2}\cos \frac{\theta _{2}}{2}\right)
^{2}+  \notag \\
&&\left. \left( \cos \frac{\theta _{1}}{2}\cos \frac{\theta _{2}}{2}\sin
\left( \phi _{1}+\phi _{2}\right) +\sin \frac{\theta _{1}}{2}\sin \frac{%
\theta _{2}}{2}\right) ^{2}\right] +  \notag \\
&&\frac{9}{4}\left( 1-p\right)  \label{payoffs-pd-a}
\end{eqnarray}%
\begin{eqnarray}
\$_{B}(\theta _{1},\phi _{1},\theta _{2},\phi _{2}) &=&p\left[ 3\left( \cos
^{2}\left( \phi _{1}+\phi _{2}\right) \cos ^{2}\frac{\theta _{1}}{2}\cos ^{2}%
\frac{\theta _{2}}{2}\right) \right. +  \notag \\
&&5\left( \cos \phi _{1}\cos \frac{\theta _{1}}{2}\sin \frac{\theta _{2}}{2}%
-\sin \phi _{2}\sin \frac{\theta _{1}}{2}\cos \frac{\theta _{2}}{2}\right)
^{2}+  \notag \\
&&\left. \left( \cos \frac{\theta _{1}}{2}\cos \frac{\theta _{2}}{2}\sin
\left( \phi _{1}+\phi _{2}\right) +\sin \frac{\theta _{1}}{2}\sin \frac{%
\theta _{2}}{2}\right) ^{2}\right] +  \notag \\
&&\frac{9}{4}\left( 1-p\right)  \label{payoffs-pd-b}
\end{eqnarray}%
For $p=1$ the above results reduce to that of Eisert et al. \cite{eisert}
and the dilemma in game is resolved for players strategies $U(\theta
_{1},\phi _{1},\theta _{2},\phi _{2})=U(0,\frac{\pi }{2},0,\frac{\pi }{2})=Q$
with $\$_{A}(0,\frac{\pi }{2},0,\frac{\pi }{2})=\$_{B}(0,\frac{\pi }{2},0,%
\frac{\pi }{2})=\left( 3,3\right) $. Next we see whether the strategy $Q$ is
NE for $p\neq 1.$ \ In this case the NE\ conditions 
\begin{eqnarray}
\$_{A}(0,\frac{\pi }{2},0,\frac{\pi }{2})-\$_{A}(\theta _{1},\phi _{1},0,%
\frac{\pi }{2}) &\geq &0  \notag \\
\$_{B}(0,\frac{\pi }{2},0,\frac{\pi }{2})-\$_{B}(0,\frac{\pi }{2},\theta
_{2},\phi _{2}) &\geq &0  \label{NE}
\end{eqnarray}%
give%
\begin{equation}
p\left( 3\sin ^{2}\frac{\theta _{1}}{2}+2\cos ^{2}\frac{\theta _{1}}{2}\cos
^{2}\phi _{1}\right) \geq 0.
\end{equation}%
The above inequality is satisfied for all values of $p\geq 0$ showing that
the strategy pair $\left( Q,Q\right) $ continues to be Nash equilibrium for
all values of $p>0.$ It shows that although state (\ref{state in}) is not
entangled for $p\leq \frac{1}{3}$ yet when shared between two players it is
proved to be a better resource as compared to classical randomness. On the
other hand at $p=0$ when the initial state becomes maximally mixed the
payoffs become $\frac{9}{4}$ irrespective of players strategies.

For CG with payoff matrix given by payoff matrix (\ref{matrix chicken}) the
payoffs given in Eq. (\ref{payoff-delta-Pi/2}) become 
\begin{eqnarray}
\$_{A}(\theta _{1},\phi _{1},\theta _{2},\phi _{2}) &=&p\left[ 3\left( \cos
^{2}\left( \phi _{1}+\phi _{2}\right) \cos ^{2}\frac{\theta _{1}}{2}\cos ^{2}%
\frac{\theta _{2}}{2}\right) \right. +  \notag \\
&&\left( \cos \phi _{1}\cos \frac{\theta _{1}}{2}\sin \frac{\theta _{2}}{2}%
-\sin \phi _{2}\sin \frac{\theta _{1}}{2}\cos \frac{\theta _{2}}{2}\right)
^{2}+  \notag \\
&&\left. 4\left( \sin \phi _{1}\cos \frac{\theta _{1}}{2}\sin \frac{\theta
_{2}}{2}-\cos \phi _{2}\sin \frac{\theta _{1}}{2}\cos \frac{\theta _{2}}{2}%
\right) ^{2}\right] +  \notag \\
&&2\left( 1-p\right)  \label{payoff-chicken-a}
\end{eqnarray}%
\begin{eqnarray}
\$_{B}(\theta _{1},\phi _{1},\theta _{2},\phi _{2}) &=&p\left[ 3\left( \cos
^{2}\left( \phi _{1}+\phi _{2}\right) \cos ^{2}\frac{\theta _{1}}{2}\cos ^{2}%
\frac{\theta _{2}}{2}\right) \right. +  \notag \\
&&\left( \sin \phi _{1}\cos \frac{\theta _{1}}{2}\sin \frac{\theta _{2}}{2}%
-\cos \phi _{2}\sin \frac{\theta _{1}}{2}\cos \frac{\theta _{2}}{2}\right)
^{2}+  \notag \\
&&\left. 4\left( \cos \phi _{1}\cos \frac{\theta _{1}}{2}\sin \frac{\theta
_{2}}{2}-\sin \phi _{2}\sin \frac{\theta _{1}}{2}\cos \frac{\theta _{2}}{2}%
\right) ^{2}\right] +  \notag \\
&&2\left( 1-p\right)  \label{payoff-chicken-b}
\end{eqnarray}%
With the help of Eqs. (\ref{NE}) the strategy pair $U(\theta _{1},\phi
_{1},\theta _{2},\phi _{2})=U(0,\frac{\pi }{2},0,\frac{\pi }{2})$ will be
NE\ of this game if 
\begin{equation}
p\left[ 2+\cos ^{2}\frac{\theta _{1}}{2}\left( 3\cos ^{2}\phi _{1}-2\right) %
\right] \geqslant 0.
\end{equation}%
The above condition is satisfied for all values of $p\geq 0.$ It means that
dilemma can be resolved in CG when the players share the state (\ref{state
in}) with $p>0$. Furthermore it can be investigated by Eqs. (\ref%
{payoff-chicken-a}, \ref{payoff-chicken-b}) that for $p=0$ the payoffs of
the players become $2$, independent of players decisions.

\textrm{Comparing our results with Fig. (\ref{fig:image}) we see that for
all values of quantum discord greater than zero there is no dilemma in both
PD and CG. Therefore it may be safe to conclude that when Werner states are
used as initial state for a quantum game it is the quantum discord the helps
resolve the dilemmas.}

For the measurement performed in product basis (i.e.,$\delta =0$ in Eqs. (%
\ref{oper 1} to \ref{oper 4}) ) the Eq. (\ref{payoffs}) reduces to 
\begin{eqnarray}
\$_{j}(\theta _{1},\phi _{1},\theta _{2},\phi _{2}) &=&\frac{p}{2}\left[
\left( \$_{00}^{j}+\$_{11}^{j}\right) \left\{ \cos ^{2}\frac{\theta _{1}}{2}%
\cos ^{2}\frac{\theta _{2}}{2}+\sin ^{2}\frac{\theta _{1}}{2}\sin ^{2}\frac{%
\theta _{2}}{2}+\right. \right.   \notag \\
&&\left. \frac{1}{2}\sin \theta _{1}\sin \theta _{2}\sin \left( \phi
_{1}+\phi _{2}\right) \right\} +\left( \$_{01}^{j}+\$_{10}^{j}\right)
\left\{ \cos ^{2}\frac{\theta _{1}}{2}\sin ^{2}\frac{\theta _{2}}{2}\right. 
\notag \\
&&+\left. \left. \sin ^{2}\frac{\theta _{1}}{2}\cos ^{2}\frac{\theta _{2}}{2}%
-\frac{1}{2}\sin \theta _{1}\sin \theta _{2}\sin \left( \phi _{1}+\phi
_{2}\right) \right\} \right] +  \notag \\
&&\frac{\left( 1-p\right) }{4}\left(
\$_{00}^{j}+\$_{01}^{j}+\$_{10}^{j}+\$_{11}^{j}\right) 
\label{payoffs-delta=0}
\end{eqnarray}%
$\allowbreak $ \textrm{For }$p>0$\textrm{\ the above payoffs remain
equivalent to the payoffs obtained by Marinatto and Weber's quantization
scheme where the players also have the option to manipulate the phase }$\phi 
$\textrm{\ of the given qubit \cite{marinatto,ma}. However at }$p=0$\textrm{%
\ when the quantum discord disappears the payoffs given by Eq. (\ref%
{payoffs-delta=0}) become average value of the entries of payoff matrix (\ref%
{payoff-matrix-general}).}

\section{\label{conclusion}Conclusion}

In this paper we quantized PD and CG by our generalized quantization scheme
taking a Werner-like state as an initial quantum state \cite{nawaz} to
explore the role of QD in quantum games. Generalized quantization scheme
allows measurements in both entangled and product basis.\emph{\ }For the
entangled basis measurement we showed that the dilemma in both PD and CG can
be resolved for all non-zero values of QD.\textrm{\ For the case of product
basis measurement the payoffs remain quantum mechanical only for nonzero
quantum discord i.e. for }$p>0.$ However \textrm{at }$p=0$\textrm{\ where
quantum discord disappears from the initial quantum state the payoffs become
constant and independent of players strategies.}

\end{document}